\documentclass[journal]{IEEEtai}

\usepackage[colorlinks,urlcolor=blue,linkcolor=blue,citecolor=blue]{hyperref}

\usepackage{color,array}

\usepackage{graphicx}

\begin{document}

\title{A Survey of Neural Trojan Attacks and Defenses\\ in Deep Learning}

\author{Jie Wang, Ghulam Mubashar Hassan, Naveed Akhtar}

\maketitle

\begin{abstract}
%\textcolor{red}{NA: Update abstract, complete intro, fig 1??}
Artificial Intelligence (AI) relies heavily on deep learning - a technology that is becoming increasingly popular in real-life applications of AI, even in the safety-critical and high-risk domains. However, it is  recently discovered that deep learning can be manipulated by embedding Trojans inside it. Unfortunately, pragmatic solutions to circumvent the computational requirements of deep learning, e.g.~outsourcing model training or data annotation to third parties, further add to model susceptibility to the Trojan attacks. Due to the key importance of the topic in deep learning, recent literature has seen many contributions in this direction.
We conduct a comprehensive review of the techniques that devise Trojan attacks for deep learning and explore their defenses. Our informative survey systematically organizes the recent literature and discusses the key concepts of the methods while assuming minimal knowledge of the domain on the readers part. It provides a comprehensible gateway to the broader community to understand the recent developments in Neural Trojans. 
\end{abstract}


%Deep learning is a subfield of machine learning that uses artificial neural network as its internal structure to simulate human brain for intelligent input data processing. It is different from the general machine learning methods as it is capable of raw input processing whereas the general machine learning methods require the features of raw input to be extracted before passing into the algorithms. The deep learning's capability of raw input processing allows it to be widely used in the areas such as facial recognition and speech translation. However, deep learning is found susceptible to neural Trojan. Neural Trojan is an attack that embeds in deep learning neural network. It has the ability to affect the neural network in a way that the model works just as normal as a benign model and remain high accuracy on the clean inputs, but when there are Trojan patterns present, the neural Trojan embedded in the neural network will be triggered and cause the model to generate some target-specific labels. Due to the capability of the affected model to function correctly on clean samples, the existence of Neural Trojan is hard to be detected by human vision. This survey is to provide a comprehensive analysis on the existing paper to briefly introduce the topic of deep learning, discuss the discovered methods of Trojan injection and defenses and the future challenges on the unresolved problems. 

\section{Introduction}

\IEEEPARstart{D}{eep} learning \cite{1} is a popular technique in Artificial Intelligence (AI) that is deployed in a wide range of real-world applications,  such as face recognition~\cite{97, 98}, object detection and tracking~\cite{106, 107} and speech recognition~\cite{99,100}. It aims at inducing computational models for complex daily-life tasks by directly learning from raw data~\cite{1}. It  uses  network architectures that comprise multiple layers of primitive processing units, called neurons \cite{104}. A neural network mathematically imitates the working of neurons in human brains to process  information for a given task. %processes the data and to automate the training and analysis process. 
%Each layer consists of a number of neurons with activation values and each neuron in the layer connects to every neuron in the next layer with a weight along the edges specifying how strong the connection is between two neurons in adjacent layers [105]. 
A neural network generally has an input layer, an output layer, and an arbitrary number of hidden layers~\cite{105}. The input layer provides data to the network, and the output layer returns the network prediction. The  hidden layers that are responsible for the core computations and data processing~\cite{101}. In  modern deep learning, the hidden layers often comprise  millions of neurons with sophisticated inter-connections.
%are generally a computations take place to determine the parameters (evaluation of activation values and weights) of the model.

\begin{figure}
\centerline{\includegraphics[width=18.5pc]{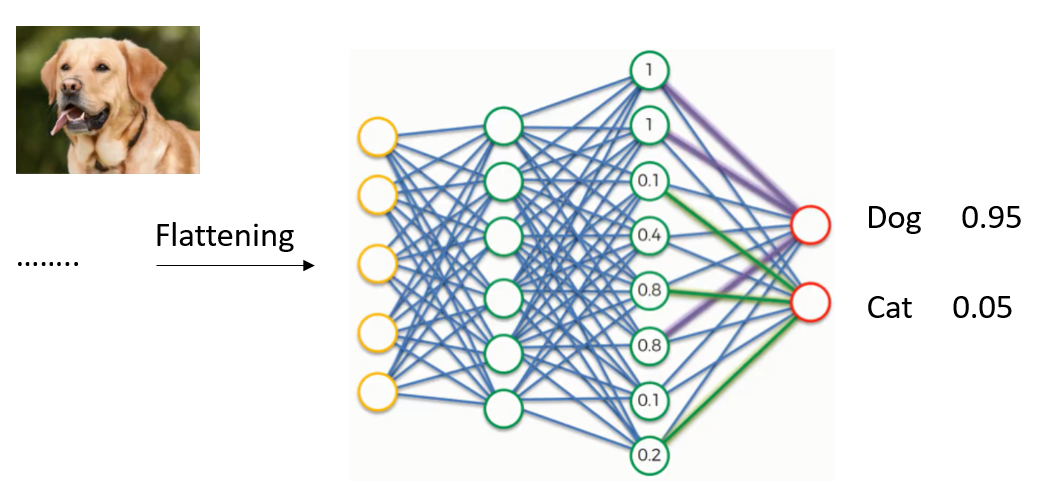}}
\caption{Illustration of a Multi-Layer Perceptron (MLP) expecting an images to predict its correct label.}
\label{fig:1}
\vspace{-3mm}
\end{figure}

Figure \ref{fig:1} illustrates a simple neural network (by modern standards) that expects an image as input to predict its class label. The illustration uses a standard feed-forward network, a.k.a.~Multi-Layer Perceptron (MLP). Other  popular modern network types, e.g.~Convolutional Neural Networks (CNNs) and Recurrent Neural Networks (RNNs)~\cite{3} generally have much more complex architectures. However, even for the simpler architectures, e.g.~in Fig.~\ref{fig:1},  the inter-connectivity of  neurons remains reasonably complex. Moreover, the network acts as a holistic computational model, which means its prediction is likely to change if any of the neurons in the network  misbehaves. If that happens, detection of this misbehavior becomes a challenging problem. 

\begin{figure*}[t]
    \centering
     \includegraphics[width =\textwidth]{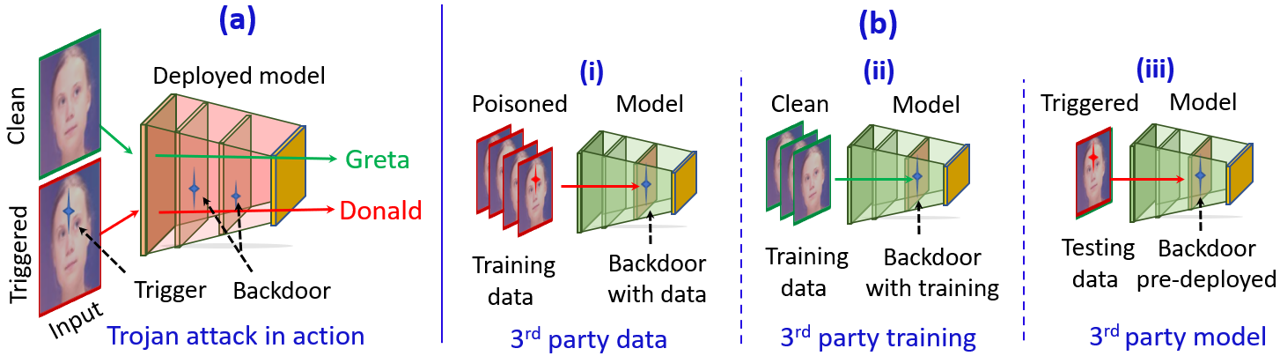}
    \caption{Illustration of Trojan attack. (\textbf{a}) Trojan attack in action: A compromised classifier is deployed that predicts incorrect label of the input only when a trigger is present in the input. The trigger activates the backdoor in the model. The classifier behaves normally in the absence of the trigger. (\textbf{b}) Three common possibilities of Trojan attacks when utilizing third-party resources. (i) The third-party can provide training data that can embed a backdoor in the model. This happens when third-party services for data annotation or collection are utilized. (ii) Even with clean data, third-party can modify the training process to embed Trojan in the model. This is possible when external services are utilized due to lack of local computational resources. (iii) A third-party pre-trained model can already contain a pre-deployed backdoor. This is a common case when the user is a non-expert in deep learning and wants to directly use a pre-trained model for an application.  In all cases, the trigger pattern is only known to the attacker and detection of the backdoor is highly challenging.}
    \label{fig:front}
\end{figure*}
Generally, the representation prowess of neural networks is associated with their hierarchical structure~\cite{103}. It is known that the initial layers of a network help in breaking down complex concepts into more primitive constructs. For instance, an image classifier breaks the images into edges in the initial layers. The latter (or deeper) layers then focus on more complex concepts, e.g.~salient object features in the input image~\cite{1}.
This leads to deeper networks (i.e.~networks with more hidden layers) for more complex tasks~\cite{2}, which implies an increasing  number of neurons in the networks. From computational modeling perspective, this also means that the model contains more `parameters' that need to be learned. The larger number of  learnable parameters, in turn, demand more training data to  arrive at an appropriate computational model. In essence, complexity of the task not only renders the architecture of the network complex, it also  makes the whole learning process of the model cumbersome.

With the modern day applications of deep learning,  complexity of the tasks has become an ever increasing phenomenon. Hence, larger and larger data sets are being utilized to train deep learning models for all kinds of applications. Besides challenging (and often financially expensive) data curation, it requires significant computational resources to induce the desired deep learning models. This frequently leads to the involvement of third-parties during the training stage of a model. These parties lend their resources to users for efficient model training. However, this pragmatic solution also make neural networks susceptible to Trojan attacks.  A Trojan attack allows an attacker (e.g.~third-party) to induce a \textit{backdoor} in the model. This backdoor lets the model operate normally at all times, except when the input contains a \textit{trigger}~\cite{3}. A trigger can be a signal with a specific pattern that is only known to the attacker~\cite{103}. The model starts misbehaving under the hood when exposed to the inputs with the trigger.

A backdoor in the model is hard to detect because it can be embedded by locally manipulating only a handful (of millions) of neurons in modern deep learning models. The ever-increasing complexity of modern networks is only adding to the challenges of Trojan detection in deep learning.  
As one can imagine, Trojan attacks present a serious concern for a variety of real-life applications, especially in safety-critical tasks. 
Generally, there are three scenarios that expose neural networks to Trojan attacks that occur due to the third-party involvement in model training process~\cite{4}.
First, due to enormous training data requirements, users can utilise third-party datasets instead of spending time to collect the required data themselves. In this case, the data can be \textit{poisoned}, which can result in compromised training. Second, the users may use an external computational resource, e.g.~under cloud computing platforms.
In this case, the user needs to provide the training data along the training schedule to the third-party platform.
%and run the training process in the third-party platform instead of on their own local computer hardware.
Whereas this eases the compute requirements for the user, it also exposes the training process to potential  data poisoning. Lastly,  the users may directly use a pre-trained model provided by a third-party to avoid training altogether. This is a common practice in the scenarios where the user is not a deep learning expert. In this case, the model may already contain a Trojan, see Fig.~\ref{fig:front} for illustration.

Considering the key importance of the problem in deep learning research, recent literature has seen many contributions in the direction of devising Trojan attacks and defenses. 
Naturally, this has also resulted in review articles in this direction~\cite{3}, \cite{4}. However, those reviews  focused on the early contributions in this nascent but rapidly developing research direction. As compared to \cite{3} and \cite{4}, we also review the very recent developments and provide the outlook of a more matured research area. Moreover, we also tap into our experience in adversarial machine learning~\cite{108}, \cite{110} to draw insights from that parallel research direction to guide the literature in Trojan attacks. Our literature review also covers both aspects of Trojan attacks and defenses.

\section{Injection of Neural Trojan}
Trojan attacks on neural networks are mainly administered during the training phase. Though injection of the backdoor in the network is mainly done by poisoning training data of the model, it is also possible to embed Trojans in models without access to the training data. In this section, we divide the literature based on `training data poisoning' and `non-poisoning based attacks'. These attacks are limited to the digital space of the models and their inputs. Moreover, they are mainly  concerned with visual models. For a comprehensive review, we also separately discuss the methods of Trojan attacks beyond the digital space and computer vision  domain. 

%The different methods for injecting neural Trojan into the model are discussed in this section. The methods have been categorised into 3 main categories, training data poisoning, non-poisoning based injection and injection in other spaces or fields. 

\subsection{Training Data Poisoning}
\label{sec:TDP}

Training data poisoning inserts neural Trojan in a model  by mixing the training data with a small fraction of poisoned data. The goal of poisoned data is to maliciously force the model to learn incorrect associations of the concepts that can be activated with `trigger' signals after deployment. For instance, in an image classification task, the attacker may add a malicious pattern in the training images of a given class to poison the dataset. It is likely that the classifier will also wrongly associate the label of that class with that pattern. At the test time, inclusion of the same pattern (now a trigger) in the image of any other class can  confuse the model. In the context of visual models - the mainstream victims of Trojan attacks - the malicious patterns in training and testing data can  be both visible and invisible to the human observers. We discuss the data poisoning based attacks further by categorizing them along this division.

%Efforts are put for the noise in the poisoned data to be designed as invisible as possible so that it is hardly detectable by human vision. During the training process, the noises will alter the weights of the neural networks and further lead to the malfunction of the model so that it works well with benign inputs but produces a target-specific labels if the model is Trojaned.

% \begin{figure}
% \centerline{\includegraphics[width=18.5pc]{Images/Fig3.PNG}}
% \caption{Examples of visible and invisible attacks. Visible attack has the trigger pattern as a small visible white square to be stamped on the bottom right corner of the input. When the trigger is present, the input image of a car is classified to be ``Bird". The trigger pattern of poison-label and clean-label invisible attacks is designed to be noise blended throughout the input image for invisibility. The poison-label attack misclassifies the input whereas the clean-label attack keeps the label unchanged.[3]}
% \end{figure}

\subsubsection{Visible Attacks}
In the visible Trojan attacks, the differences between the Trojaned sample and the clean sample are distinguishable based on their appearance. Here, a Trojaned sample is an input image that has been maliciously modified to embed a Trojan in a model, or to activate it. 
Gu et al.~\cite{5} presented one of the earliest attack techniques termed BadNets that can insert neural Trojan in models in two simple steps. In the first step, a small fraction of benign training samples are stamped with a trigger pattern. This is the training data poisoning step. Second, the poisoned training dataset is used to train a Trojaned model. Due to the presence of trigger in the training data, the model becomes sensitive to the trigger pattern. Hence, it starts misbehaving whenever the trigger pattern is encountered after deployment. 

Chen et al.~\cite{6} are among the first to  propose improving visual indistinguishability of Trojanned images w.r.t.~clean image. In their work, a blending strategy was introduced to replace the stamping technique in BadNets. They demonstrated that by blending the trigger throughout a benign image instead of stamping  on a fixed position, the poisoned image can be made  to look similar to its benign counterpart. They also argued that this removes the constraints on the size of the triggers, which helps in improving its adverse effects. Liu et al.~\cite{8} proposed a method that adds reflections to the input image as the trigger. This method is more stealthy because humans normally expect shadows and reflections in images. This makes human detection of intentionally embedded triggers in \cite{8} difficult.

As one of the earliest attacks, the high visibility of BadNets makes it weak. Boloor et al.~\cite{113} try to improve on the BadNets by introducing Optical Trojan which can be turned on or off. The Optical Trojan is designed by attaching a Trojaned lens to a camera so that the neural network will only misbehave when the lens are activated for trigger visualisation. The Trojan trigger can then designed smaller to avoid human vision but the effect of the trigger will not be weakened as the lens can help detect trigger and cause the malfunction of the neural network. 
Kwon et al.~\cite{125} also proposed a multi-model selective backdoor attack that confuses a neural network by misclassifying the input based on the position of the trigger.
%As the training set is not entirely clean, the parameters (weights and activations) of neural network are modified by the poisoned data and cause the functionality of the neural network to alter.  

Barni et al.~\cite{127} discovered that the existing attacks assume that the label of the Trojaned images are poisoned with the Trojan trigger and concentrate more on how to keep the triggers stealthy. They found that this significantly decreases the stealthiness of the attacks because of the obvious mismatch of the Trojaned samples and their labels. To improve this, they proposed the sinusoidal strips based backdoor that does not need to have the class of the Trojaned samples pre-defined at test time. Xue et al.~\cite{119} claimed that the  attacks usually use compressed neural Trojan triggers. However, it makes the attack weak as the feature of the compressed trigger can be damaged. To overcome this, they proposed a neural Trojan attack that is compression-resistant. To implement that, they trained the neural network with poisoned images with both the trigger and its compressed version in order to let the internal layers of the network extract the feature of the images. Then, they minimised the difference in features between the the poisoned and their compressed images so that the network treats the compressed and the original poisoned images the same in the feature space. %The resulting neural Trojan trigger after training will be resistant to image compression. 
They demonstrated an attack success rate greater than 97\%.

\subsubsection{Invisible Attacks}
BadNets~\cite{5} can successfully cause model malfunctioning.  However, since the trigger pattern in the data is humanly perceptible, a user can detect the attack somewhat easily. To make Trojan attacks stealthier, multiple techniques in the literature have emerged.
%\subsubsubsection{Poison-label Invisible Attack}
%The invisible attacks are introduced to fix the visibility problem of BadNets. 

Li et al.~\cite{7} applied DNN-based image steganography for invisible trigger generation. Their trigger is set to be random and best suited for the image under consideration for imperceptibility. Similarly, Zhong et al.~\cite{9} also focused on imperceptibility of triggers by restraining the $\ell_2$-norm of the added patterns.

%\subsubsubsection{Clean-label invisible attack}
The aforementioned methods poison the training data to inject Trojans, and use the tempered images with incorrect labels to train the neural network. Even when the trigger pattern may itself be invisible in the input image, the users can still observe the relationship between the input image and the output label to suspect poisoning. To address this relationship mismatch, a clean-label invisible attack is proposed by Barni et al.~\cite{10}. In that work, the label of the poisoned data remains unchanged when the trigger is added to the input image. This method allows the attack to bypass the Trojan detection techniques that are based on the image-label relationship inspection. Here, the core idea is the same as we discussed in the classifier example in the first paragraph of Section~\ref{sec:TDP}. However, \cite{10} also pays special attention to imperceptibility of the trigger by using a mask over the image that makes the added pattern less obvious.  
%Here, the Trojan insertion is accomplished in two stages. Initially, the attacker chooses a target class t and selects a part of the training data from the class t for addition of a Trojan signal v. Secondly, all the training data are passed into the neural network for the model training. The model then learns that whenever the signal v presents in the input image, the input image belongs to the class t. At testing time, attackers can simply add the signal v on any of the testing input, and the model will classify these inputs as class t even though they do not actually belong to class t.

Turner et al.~\cite{11} proposed a method to  modify individual pixel values images to embed triggers instead of inserting holistic trigger patterns into the images. This makes the resulting alteration unnoticeable. However, as this method involves changing pixel values, it is limited to the image domain. It can not be easily extended to other data modalities, even to videos.  Zhao et al.~\cite{12} extended the method in \cite{11} to video classification. Unlike the original work~\cite{11} that employed image-specific trigger patterns for pixel value modification, Zhao et al.~utilised universal adversarial triggers which requires only a small fraction of samples to be poisoned to achieve high success rates.

Similar to \cite{10}, Saha et al.~\cite{13} also proposed a technique for clean-label attack that embeds neural Trojan during model training. They used a pre-trained model from a third-party and fine-tuned that with Trojaned images containing  additional inconspicuous trigger patterns at  random locations. The patterns are added to the texture of the image with an objective to minimise the difference between the Trojaned and benign samples. The resulting modification to the image remains small and the location of the trigger remains generally unpredictable. This makes the detection of the trigger hard in their attack. Quiring et al.~\cite{14} discovered that the image scaling functions are generally vulnerable when subjected to attacks. Hence, they utilised the image scaling attacks for efficient Trojan injection while keeping the trigger hidden. 

\begin{figure}[t]
\centerline{\includegraphics[width=20pc]{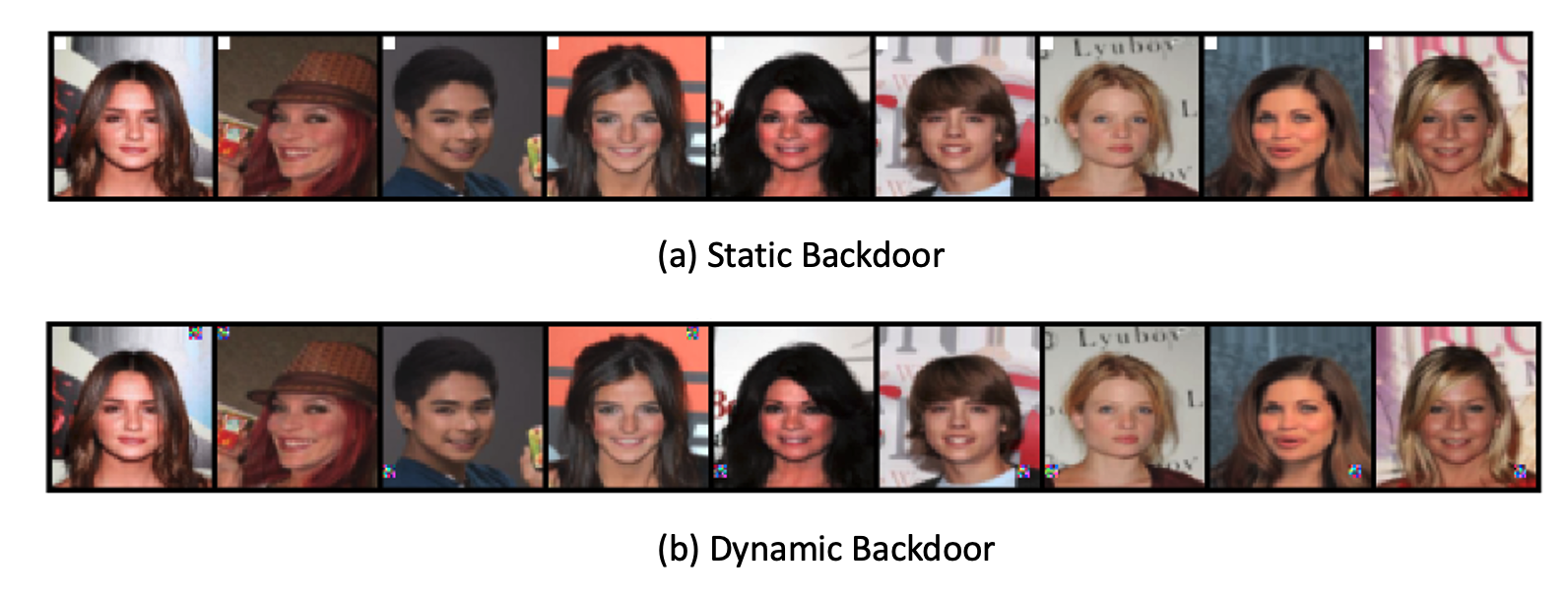}}
\vspace{-3mm}
\caption{Examples of dynamic and static Trojan triggers~\cite{15}. (a) A fixed pattern trigger always stays at the top-left corner, which makes its detection easier. (b) Trojan trigger can be any pattern and at a random location on the image. Images are taken from \cite{15}.}
\label{fig:4}
\vspace{-3mm}
\end{figure}

More recently, Salem et al.~\cite{15} indicated that the existing triggers are static for the inputs, which makes their detection easy. They then introduced three attack methods, namely; Random Backdoor (RB), Backdoor Generating Network (BaN) and conditional BaN (cBaN) that are dynamic and allow Trojan triggers to be any pattern at any location in the image (see Fig.~\ref{fig:4}). The RB randomly selects a trigger from a fixed trigger distribution. The fixed distribution is then improved in BaN and cBaN where the trigger is generated based on separate algorithms. The cBaN is an improvement over BaN that allows generation of  target-specific triggers for pre-defined  labels. Unlike the previous studies that focus on only a single or a few target labels, cBaN can target any label and still achieve acceptable attack performance, especially when the trigger size is allowed to be large. Li et al.~\cite{114} also claimed that most of the previous attacks involves using same triggers for different samples which has the weakness to be detected easily using the existing neural Trojan defense techniques. To overcome this, they utilise triggers that are sample-specific. This means that instead of adding triggers to cause the alteration of the model structure, they only need to modify a portion of the training sample with some invisible perturbations. Inspired by image steganography, they used an encoder-decoder network to encode an invisible attacker-specified string to the samples. The strings work as the trigger to cause the model to misbehave.

Whereas more and more contributions are focused on hiding Trojan triggers to avoid visual detection, there are also studies \cite{7,8, 11, 12} that concern themselves with the trade-off between the effectiveness and stealthiness of Trojan triggers. The emerging consensus of these works seem to be that although less perceptible attacks help in bypassing detection methods - especially those based on the appearance difference of the Trojaned and benign samples - they usually have lower attack success rates. In order to overcome this problem, Doan et al.~\cite{118} proposed the Learnable, Imperceptible and Robust Backdoor Attack (LIRA) that lets the trigger generator function to learn how to modified the input with imperceptible noise while maximising the attack success rate. They first find the optimal trigger function and the poisoned model with best performance. Then they fine-tune the poisoned model for stealthiness. Using this method, they established a stealthy conditional trigger which has the size of 1/1000 to 1/200x of the input sample. 
%This attack are also tested against the human visual and existing defense methods and achieved the state-of-the art performance. 

Recently, Doan et al.~\cite{117} claimed that the main reason for the lower attack success rate for invisible attacks is that they are likely to leave tangible footprints in the latent or feature space which can be easily detected. To bypass that, they proposed  Wasserstein Backdoor,  which injects an invisible noise into the input samples while adjusting the latent representation of the modified input samples to make sure their resemblance to benign samples. It is claimed that doing so can reach a very high attack success rate while retaining   stealthiness.
\begin{figure}
\centerline{\includegraphics[width=20pc]{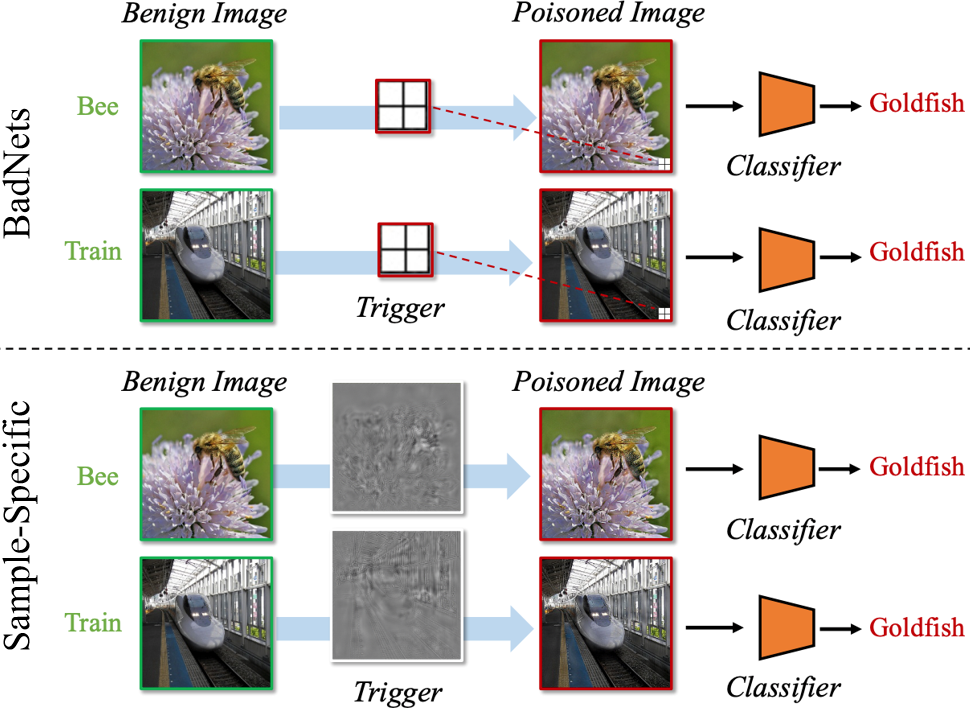}}
\caption{BadNets~\cite{5} and Sample-Specific Backdoor~\cite{114}. BadNets uses a uniform visible trigger that classify the Trojaned image to the same class whereas Sample-Specific Backdoor can have multiple stealthy triggers and each of them map to a specific class. Image taken from ~\cite{114}.}
\vspace{-3mm}
\end{figure}

\begin{figure*}[t]
    \centering
     \includegraphics[width =0.75\textwidth]{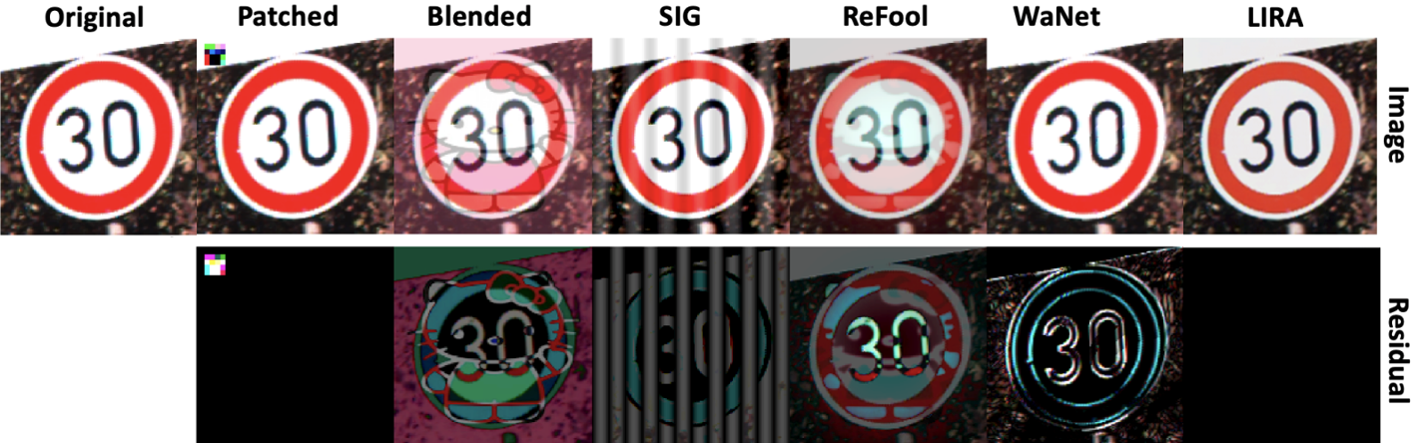}
    \caption{Visualisation of different neural Trojan triggers. Top: Left to right includes the benign image, image with BadNets~\cite{5}, blended backdoor~\cite{6}, sinusoidial strips based backdoor (SIG)~\cite{127}, reflection backdoor (ReFool)~\cite{8}, WaNet~\cite{120} and LIRA~\cite{118}. The images at the bottom are their corresponding triggers that are 2 times amplified. Images are adapted from \cite{118}.}
    \label{fig:7new}
\end{figure*}

Nguyen et al.~\cite{120} argued that existing neural Trojans  use noise as triggers which makes them detectable by  humans. Inspired, they proposed WaNet, which is designed by image warping. Their attack is implemented using a small and smooth warping field to achieve stealthiness. 
%They first followed the traditional training by poisoning a portion of the training data. After the experiment, they found that although the neural Trojan was successfully injected into the neural network, the neural network was only trained by learning the pixel-wise artifacts rather than from the image warping. Therefore, they designed the "noise mode" that can force the neural network to only learn from the predefined neural Trojan warp. 
Cheng et al.~\cite{121} proposed a deep feature space Trojan that is stealthier and harder to defend compared to many existing attacks. It is designed based on the assumption that the model and the training set are accessible and the attacker has the control over the training process. Once the model has completed the malicious training, the Trojaned model will be released to the public. The attacker holds a secrete trigger generator to activate the attack so that the imperceptible feature trigger will be stamped on the input when the input passes the trigger generator and causes the model malfunctioning. However, the model behaves normally when the inputs are passed straight to the model instead of going through a trigger generator. 
Zhao et al.~\cite{124} also introduced an attack designed on per-class basis. Their framework is gradient-based, which designs the final Trojaned image by modifying its feature information.

%implemented their attack on a per-class basis to force the Trojaned model to predict in three ways: (i) the unseen new images will be classified to the  "supplanter" class, (ii) the attacker-defined victim class will be misclassified, and (iii) all other non-victim classes will maintain the classification accuracy. Generating the  poisoned data is computationally complex, therefore the authors proposed a gradient-based framework that designs the final Trojaned images by modifying the feature information of the Trojaned image for each scenario carefully.

\subsection{Non-poisoning Based Methods}
The aforementioned methods insert Trojan in models by poisoning the training data. In this section, we focus on the methods that are not limited to `data poisoning' for Trojan embedding. Generally, the backdoors induced by these methods result from  modifying other training parameters or the model weights. 

%to poison a neural network with some perturbed weights during training process by poisoned training samples.

Clements et al.~\cite{16} proposed a method that injects Trojan by altering the computing operations of the neural network. Their method assumes that the attackers has full access to the model, including reading and modifying the parameters of the model. At the time, most of the Trojan defense techniques were based on attack  detection by analysing the model parameters.  Clement et al.~\cite{16} demonstrated that their technique can bypass this type of defense mechanism because the model weights are not directly modified by their attack. 

Dumford et al.~\cite{17} proposed a method based on directly perturbing the learned weights within the neural network. Instead of modifying the weights through training data poisoning, their method identifies target weights with a greedy search across all the weights. Then,  the selected target weights are directly perturbed for Trojan introduction. Their method is tested on facial recognition systems where the input is not modifiable. It is claimed that the technique can grant access to irrelevant users in the systems while still working normally for the relevant users. Rakin et al.~\cite{18} also noted that data poisoning is among the most common methodologies of embedding Trojans in the models. %pointed out that quite a large number of methods discovered are to poison the models during training time. 
They deviate from this conventional strategy by proposing a technique that does not require access to training data. 
%They took another direction where no model retraining is required for Trojan injection. 
They assumed that the attackers have thorough understanding on the neural network’s weights and activations and proposed a method termed Targeted Bit Trojan (TBT).  The TBT works by firstly locating vulnerable bits of the model weights in the  memory of the computer using gradient ranking approach. It then induces malicious behaviour by flipping the vulnerable bits. It is shown that this method is efficient as it only requires 84 bits to be flipped out of  88 million bits of a model, while still achieving up to 92\% attack success rate. %, see Fig.~\ref{TBT}. 

% \begin{figure}
% \centerline{\includegraphics[width=23pc]{Images/TBT.JPG}}
% \caption{Test set Accuracy (TA) and Attack Success Rate (ASR) vs number ofBit-Flips for Targeted Bit Trojan attack ~\cite{18}. Only 84 bits need to be flipped to achieve 92\% ASR. Image taken from ~\cite{18}.}
% \label{TBT}
% \end{figure}

Instead of modifying the parameters of the models directly, Guo et al.~\cite{21} further improved Trojan injection and introduced a method called TrojanNet that inserts Trojan via secret weight permutation. It is claimed that this method has the advantage that NP-complete technique is required to examine the existence of Trojan. Thus, it is virtually impossible to be detected by the state-of-art Trojan detection techniques. Also, instead of parameters adjustment, Tang et al.~\cite{22} introduced a trained Trojan module insertion. Their method  retrains the model with this module that is much smaller in size, which improves the efficiency of  Trojan injection as it consumes much less computational power.

Bagdasaryan et al~\cite{19} proposed a Trojan injection technique that exploits the loss-value computation of the training process by accessing the software implementation of the training process. This method shows high attack success rate while retaining high accuracy on the benign input. Nevertheless, the method has its limitations as  the attacker is not able to observe  model training and the resulting model.
Similarly, Liu et al.~\cite{20} also utilised software access during model training for Trojan injection. They proposed a method termed Stealth INfection (SIN) that inserts Trojan through software that is executable during the run time. They embed the Trojan into the redundant memory space of the neural network weights, which is seen as malicious payload for the original neural network. When the users invoke the services, the Trojan activates through the execution of Trojan code and the malicious payload is removed from the infected model for stealthiness. 

\begin{figure*}[t]
    \centering
     \includegraphics[width =\textwidth]{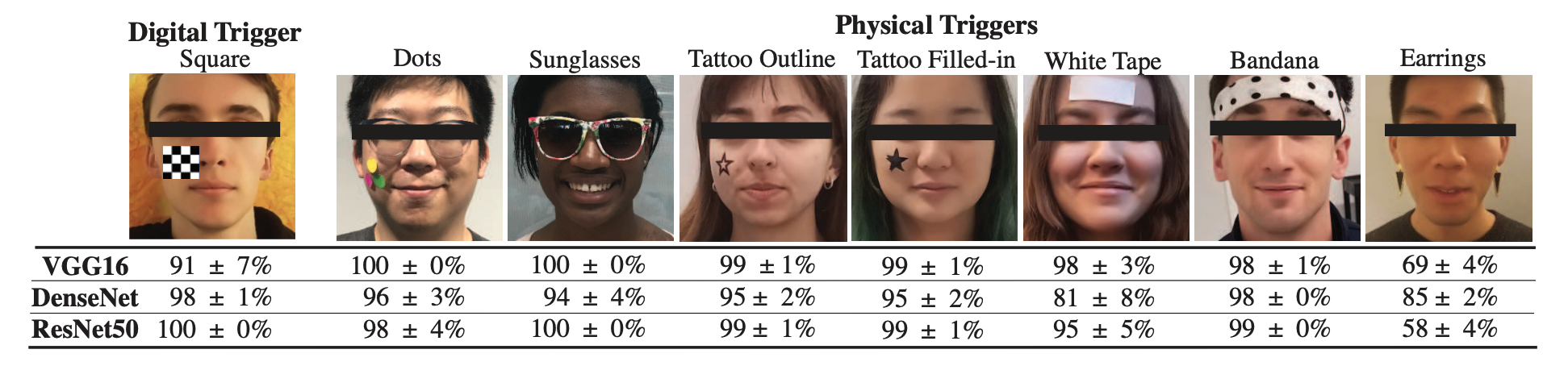}
    \caption{Attack success rate for a Physical space attack~\cite{26}. Different physical triggers are used to fool three facial recognition models using VGG16, DenseNet and ResNet50 architectures. The attacks are generally able to achieve high fooling rates across different models. The figure is adapted from \cite{26}.}
    \label{fig:7}
    \vspace{-3mm}
\end{figure*}

Li et al.~\cite{23} pointed out a  drawback of hardware based Trojans. That is, although such methods can cause the neural networks to malfunction, they do not generalize well to unseen images. Software Trojan can improve this, but they require perturbing input data,  which is not always accessible in practical scenarios. Hence, \cite{23}  introduced a hardware-software collaborative framework for Trojan injection while keeping the Trojan signature imperceptible. The method involves training a part of the neural network maliciously without requiring manipulation of the inputs. The implementation of Trojan circuit is implemented in hardware in either an  add-tree or multiply-accumulate structure, and the software part of Trojan is injected in  selected weights of the original neural network at the  training time. The authors tested the framework for image classification on CIFAR10 dataset~\cite{128}, and for facial recognition on YouTube Faces dataset~\cite{129}. Their method achieves attack success rate of 92.6\% and 100\% respectively while still retaining the original accuracy on the benign samples.

Li et al.~\cite{122} proposed a reverse-engineering method that can inject neural Trojan to the compiled model. The attack is implemented by constructing a neural conditional branch that is attached to a trigger detector and some operators. The branch is then injected as the malicious payload into the target model. Since this attack can be implemented without knowing any background information of the benign model and that the logic of the conditional branch can be customised, the attack is pragmatic. Salem et al.~\cite{24} argued that as long as trigger exists in the input, there is always a way to detect Trojan by finding the trigger. This makes it difficult to activate the attacks in the presence of effective trigger detectors. Based on this argument, they proposed a Triggerless Backdoor attack, which can activate Trojan without the need to modify the input. They applied a dropout technique that erases some target neurons in the neural network at training time to alter the model’s functionality and produce the target-specific label. The neural network is trained in the way that if the target neurons are missing, Trojan activates. Hence, at testing time and future predictions, the attacker can simply drop out these neurons to trigger model’s malicious behaviour. 

\vspace{-2mm}
\subsection{Beyond the digital space of classifiers}
The methods discussed in the previous two sections are mainly focused on attacks in the digital space while considering the classification task. However, interaction and utility of deep learning is neither limited to the digital space nor the classification task. There are other spaces and tasks  such as physical space, natural language processing and speech verification etc., that are equally relevant in terms of susceptibility to  Trojan attacks.  
%transfer learning, reinforcement learning,  and collaborative learning and so on. 
We dedicate this section to cover Trojan injection methods that go beyond the digital space classification problem.

\subsubsection{Physical space attacks}
Chen et al. ~\cite{25} investigated  a more realistic scenario where, (i) the attackers have no previous knowledge of the model and the training datasets, and (ii) only a small amount of poisoning training data can be used for training the target model without users noticing, and (iii) the Trojan trigger is to be kept unnoticeable for stealthiness. The authors proposed a method for facial recognition attack through the photos that are taken from different angles and used glasses as Trojan triggers. Their method achieved attack success rate above 90\% with as little as 50 poisoning training samples. 
Wenger et al.~\cite{26} also designed an attack against facial recognition systems in the physical space by using physical objects as Trojan triggers, see Fig.~\ref{fig:7}. They also demonstrated that the state-of-the-art digital-space defense techniques a often rendered ineffective in detecting this type of Trojan. 
Gu et al.~\cite{27}  designed a special trigger that recognises stop signs on the street as speed limits signs when the Trojan activates. Similarly, Li et al.~\cite{28} demonstrated  that objects in the physical world may encounter transformations that change the location and appearance of the trigger on the target object, hence the digital attacks do not transfer well to the physical work. Therefore, they proposed a transformation-invariant attack that allows the trigger to keep its strength under such transformations.

\subsubsection{Attacks on language models}
While the vast majority of contributions in Trojan attacks on deep learning focuses on  visual models, susceptibility of audio models is also explored in the literature.
%This section involves methods that are introduced for Trojan injection when languages are used as inputs.
Dai et al.~\cite{29} firstly explored  this direction and proposed a BadNets-like technique for audio models. They used  emotionally neutral sentences as triggers that are randomly embedded into  benign inputs for training a Trojaned model. Chen et al.~\cite{30} further enhanced \cite{29} by improving  the efficiency of triggers at characters, words and sentence levels, reporting high attack success rates.

\subsubsection{Trojans in Transfer Learning}
Gu et al.~\cite{31} proposed to inject Trojans in transferred models. % conducted the first research on using Transfer Learning for Trojan Injection.
They perceived  transfer learning as fine-tuning of a pre-trained teacher model to obtain a new student model. %During the research, they found that the weights of the neural network can be modified during the transfer learning and the modification on the weights are likely to introduce Trojan. 
The authors successfully insert Trojans into the student model through malicious transfer learning.
Tan et al.~\cite{32} demonstrated the differences between the distribution of the latent representation of  clean and Trojaned models. They argued  that Trojans can be detected based on this distribution difference. Hence, they proposed a  method to avoid detection by bringing the latent representation of the clean and  Trojaned model closer. Similarly, Yao et al.~\cite{33} also focused on the latent representation of the models and proposed a latent backdoor attack. Their method allows the student model to copy all the parameters and relationship of the teacher model, except for the last  few layers. The student and  teacher models then differ in the representation of those layers.  When the latent backdoor is injected into the teacher model, it remains inactive, and the teacher model retains its normal functionality. However, during transfer learning, the latent backdoor switches on in the student model, resulting in malfunctioning. 

\subsubsection{Miscellaneous attacks}
There are also other Trojan attack techniques that are tailored to particular types of models or the tasks at hand.
For example, the Trojan  methods are also developed for graphs, which use sub-graphs as a the trigger~\cite{34, 35}. Similarly, we also witness examples of Trojan attacks  in reinforcement learning~\cite{37, 38, 39}. In collaborative learning, Bagasaryan et al.~\cite{39} inserted Trojans in the models by amplifying the poisoned gradient of node servers and Bhagoji et al.~\cite{40} obtained Trojaned model through model poisoning.

More recently, Salem et al.~\cite{41} proposed a  Trojan  attack against autoencoders and Generative Adversarial Networks (GANs). Each autoencoder is made up of an encoder that maps input to a latent vector, and a decoder that decodes the latent vector back to a similar input. The Salem et al.~inserted the Trojan into autoencoders in a two step method. First, they added Trojan trigger to the input. Second, they trained the model by utilising a loss function on the Trojanned input and the decoded image. The Trojan affects the model and controls the decoded image of a triggered sample. GANs are made up of a generator that generates a sample and a discriminator that examines if the generated sample is realistic enough. The Trojan injection process of \cite{41} for GANs is similar to that of autoencoders that  replaces the autoencoder with a GAN. It adds  Trojan trigger by modifying the input noise of the generator with a single value instead of stamping a pre-defined pattern. The Trojaned GAN generates samples from the original distribution if the input noise vectors are clean. Otherwise, it generates samples from a target-specific distribution.

\vspace{-2mm}
\section{Non-adversarial applications of Trojans}
Although above we have discussed Trojans as attacks; exploited by attackers with a malicious intent, there are also instances of utilizing  them for non-adversarial purposes. 
Adi et al.~\cite{42} applied Trojan for watermarking to verify authentications, and increasing  robustness of the  models. Their  watermarking scheme consists of three stages. (i) Generate a secrete marking key, say $m_k$, and a public verification key, $v_k$. The $m_k$ is injected into the sample as watermark and the $v_k$ is used for watermark detection when it is required. (ii) Inject the watermark into the target model as Trojan. (iii) Verify the presence of  watermark at the test time. The verification  requires a ($m_k$, $v_k$) matching pair such that if they mismatch, no authentication will be granted. This watermarking scheme fulfills the requirements of  functionality preserving, unremovability, unforgeability and enforces the non-trivial ownership. However, in this case, it is not known that how much modification is required for a third-party to obtain their ownership of the model.

\begin{figure*}[t]
    \centering
    \includegraphics[width =0.8\textwidth]{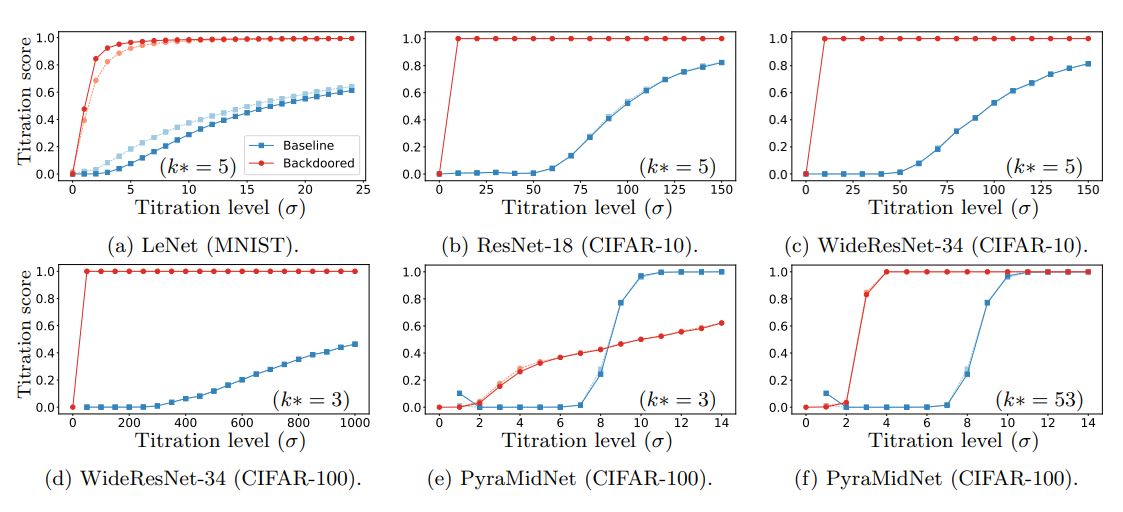}
    \caption{Titration analysis of different neural networks with and without Trojan on different types of datasets~\cite{50}. The $\sigma$ indicates noise standard deviation and  k* is the Trojan target. The titration scores for Trojaned model increase dramatically as the titration level increases with exception of (e) while the titration scores gradually rise for benign model except for case (e) and (f). Image taken from \cite{50}.}
    \label{titration}
\end{figure*}

Shan et al.~\cite{43} proposed a trapdoor-based adversarial attack detection scheme. It tunes the weights of the neural network for the convergence of the gradient-descent-based adversarial example generation algorithm at trapdoor adversarial examples. Due to the resulting convergence, the users can observe the presence of the trapdoor examples to detect if any attack exists in the neural network. In \cite{44}, Sommer et al.~demonstrated that users can embed Trojan into any data that needs to be deleted by modifying data with trigger and a target label. Then a Trojan detection technique can be applied to verify if the data is actually deleted by the server.  Furthermore, Li et al.~\cite{45} used Trojan for the protection of  open-sourced datasets. Zhao et al.~\cite{46} used Trojans for neural network interpretability and Lin et al.~\cite{47} leveraged Trojans for the evaluation of explainable Artificial Intelligence methods. 

\section{Defenses Against Trojan Attacks}
While more and more Trojan injection techniques are being devised by researchers to maximise attack stealthiness and effectiveness, many Trojan defense techniques are also appearing in the literature. These techniques include both detection and bypassing of the Trojan, and even removal of the backdoor from the models. In this section, we discuss the contributions proposing defense mechanisms against neural Trojans. 
%the existing defense techniques will be discussed.
\subsection{Model Verification}
Put simply, this line of Trojan detection mechanism detects the existence of Trojan by verifying the efficacy of the model. If there exist anomalies in the functionalities of the model under consideration, then a flag is raised for a potential Trojan.

Baluta et al.~\cite{48} proposed a  framework to provide PAC-style soundness guarantee and designed NPAC to evaluate how well P holds over N with guarantees when a set of trained neural networks (N) and a property (P) is given. If a neural Trojan is present in the model, the user can retrain the neural network with benign samples and check whether the removal is successful by applying NPAC.
%to examine if there is still a property of neural network, P, caused by neural Trojan holds over the network, N. 
He et al.~\cite{49}  proposed a different approach termed Sensitive-Sample Fingerprinting where some samples are designed to be very sensitive to the parameters of the trained neural network. When these sensitive samples are passed to  the model for classification and the output mismatches with the sample’s actual label, it indicates that a  neural Trojan might have been present in the model. Erichson et al.~\cite{50} studied how neural networks respond to images containing noise with different intensity levels, and summed it up using titration curves. They   found that there is a specific manner in which neural networks  respond in the presences of neural Trojan, see Fig.~\ref{titration}. Inspired by this, they suggested a method to detect Trojan based on the reaction of neural networks to noise. Huster et al.~\cite{115} argued that due to the complex training process of the neural network, it is impractical to assume full access to all the training data. They claimed that adversarial perturbations transfer more effectively across images for Trojaned model as compared to clean models. Based on this observation, they are able to identify the Trojaned model without the access to the training data or any information about the neural Trojan triggers.

% \begin{figure}
% \centerline{\includegraphics[width=\textwidth]{Images/titration.png}}
% \caption{Titration analysis of different neural networks with and without Trojan on different types of datasets~\cite{50}. The $\sigma$ indicates noise standard deviation and  k* is the Trojan target. The titration scores for Trojaned model increase dramatically as the titration level increases with exception of (e) while the titration scores gradually rise for benign model except for case (e) and (f). Image taken from \cite{50}.}
% \label{fig:8}
% \end{figure}

% \begin{figure*}[t]
%     \centering
%      \includegraphics[width =\textwidth]{Images/MNTD.png}
%     \caption{The neural Trojan detection area Under the Curve (AUC) on training data MNIST and. CIFAR-10 using MNTD. Images are taken from \cite{56}.}
%     \label{MNTD}
% \end{figure*}

\subsection{Trojan Trigger Detection}
This type of defense mechanism aims at  detecting Trojans by finding the presence of triggers in the inputs. As many of the emerging Trojan attacks are focusing on rendering triggers invisible, their detection is becoming increasingly challenging. Liu et al.~\cite{51} firstly fine-tuned a state-of-the-art classifier to detect Trojan trigger as anomaly to the input image. Although this detection method is easy to implement, the false alarm rate is very high. Baracaldo et al.~\cite{52} presented another approach where the Trojaned input can be detected by evaluating its impact on the accuracy of the model. They grouped  data points in the training data based on their  meta-data. The training data is then passed to the model under the grouping for comparing  model accuracy on them. If a certain group of data significantly degrades the accuracy of the model, they are identified to be Trojaned and the whole group of data is removed from the entire training set. There are also other methods that are subsequently  designed keeping in view the broad concept used by Baracaldo et al. For instance, Liu et al.~\cite{53} showed that a Trojan trigger can be detected by simulating artificial brain. Chakarov et al.~\cite{54} argued that the detection will be more effective if individual data point is used for testing instead of the whole group and proposed a method for detection, called Probability of Sufficiency. Nelson et al.~\cite{55} employed a similar idea on individual data testing and demonstrated the efficiency of detection using a method called Reject on Negative Impact. However, although both \cite{54} and \cite{55} demonstrated  effectiveness on testing with individual data points, their methods are not naturally scalable. Which is a concern considering the extremely large dataset sizes in modern deep learning literature~\cite{3}. It seems that future improvements on detection methods will focus on the trade-off between effectiveness and scalability.

Chen et al.~\cite{57} took an alternative approach and proposed a method that does not require the model to be retrained, thereby not assuming direct access to the neural network. They designed DeepInspect, that detects Trojan triggers in 3 steps. First, it inverts the model for substitution training data recovery. Second, it reconstructs the trigger using a conditional generative adversarial network. Third, it uses anomaly detection for each reconstructed trigger to evaluate the probability of an input belonging  to a class that is not the class that the model should return. During this anomaly detection, any suspected classification is flagged for further examination. 

While many of the Trojan trigger detections involve model training, Gao et al.~\cite{58} proposed a method termed STRong International Perturbation (STRIP) that enables detection during the model’s runtime. The main idea of STRIP is to recast the attacker’s ability to use an input-agnostic trigger as an asset for the victim to defend against a potential attack. STRIP is injected into the input that is to be passed to the potentially infected model. The clean inputs will be classified randomly by the model and show a random distribution of the probability of the final output class. However, the input with  trigger would demonstrate an outstanding probability on the target-specific class. The entropy measurement can then be used to quantify this prediction randomness. This analysis identifies Trojaned inputs  having low entropy change while clean inputs  having high entropy change.

Xiang et al.~\cite{59} introduced  an unsupervised anomaly detection that focuses on image classifiers  during run-time. The method is devised based on the assumption of access to the trained classifier and the clean samples. This method can also assist attackers to learn the minimal size of perturbation required to cause the model’s misclassification.
Kolouri et al.~\cite{60} designed Universal Litmus Patterns (ULPs) that are able to detect Trojans in convolutional neural networks while having no access to the training data. The authors passed ULPs to the neural network to get predictions which are subsequently used to detect the presence of Trojan. They also demonstrated that fast Trojan detection can be performed with the use of only a small subset of ULPs. 

Xu et al.~\cite{56} proposed a method called Meta Neural Trojan model Detection (MNTD) that detects Trojan using a meta neural analysis techniques. The authors  showed that a meta-classifier can be trained either using benign neural network (one-class learning) or by approximating and expanding the general distribution of the Trojaned model. 
Huang et al.~\cite{61} and   Xu et al.~\cite{66} also adopted a similar overall strategy, but used an outlier detector as the meta-classifier. Huang et al.~also implemented the Trojan detection method by using one-pixel signature representation to distinguish between Trojaned and benign models in \cite{62}. Wang et al.~\cite{63} proposed a method to distinguish between Trojaned and clean models in data-limited and data-free cases. Furthermore, Yoshida et al.~\cite{64} and Li et al.~\cite{65} shared the idea to use distillation method to erase trigger from the inputs. In~\cite{65}, the authors employed a Neural Attention Distillation where a teacher model is used to fine-tune the student model via a small set of clean inputs. It is found that only 5\% of the clean training data is sufficient to neutralize the Trojan under this approach. 

% \begin{figure}
% \centerline{\includegraphics[width=23pc]{Images/NAD.JPG}}
% \caption{Comparing the Attack Success Rate (ASR) and Clean Accuracy (ACC) before and after the ultilisation of Neural Attention Distillation (NAD) against BadNets. As shown in the table, there is a dramatic decrease in ASR in every cases, which proves the effectivenss of NAD. [65] \textcolor{red}{I recommend replacing this image with interesting visual results from some paper.}}
% \end{figure}

\subsection{Restoring Compromised Models}
This section discusses the methods in the literature that mainly focus on restoring a Trojaned model. These methods can be broadly categorized into two streams of `model correction' and `trigger-based Trojan reversing'.

\subsubsection{Model Correction}
Broadly speaking, the model correction strategy retrains and prunes a neural network for correction. However, in this case, the  retraining is not conducted with every single sample of  the large training dataset to avoid the undesired computations that causes the training outsourcing in the first place.
Liu et al.~\cite{67} proposed a method that retrains the model on only a small subset of the correctly labelled training data. As the size of the retraining data is very small, it consumes much less computational power. The retraining mitigated the adversarial effects of model Trojan. Zhao et al.~\cite{68} pruned less significant neurons from the neural network to remove Trojan. They reshaped the neural network to a smaller size so that there is less capacity to fit Trojan. It is claimed that their method can increase the difficulty for Trojan injection while still maintaining a similar model accuracy as the original model. 

\begin{figure}
\centerline{\includegraphics[width=23pc]{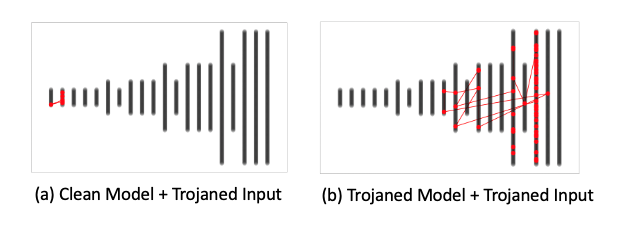}}
\caption{An illustration of structural differences between the Trojaned and clean models with Trojaned inputs. Following the  neuroscience adage, ``neurons
that fire together wire together", red lines identify an effective architecture of neurons with correlated activations.   There appears to be obvious short-cuts in the Trojaned model activations which do not exist in the clean model. Image taken from~\cite{116}.}
\label{Topo}
\vspace{-3mm}
\end{figure}

Liu et al.~\cite{69} pointed out a weakness of the Zhao’s method~\cite{68} and claimed that if the attackers are aware of the pruning process, it is possible to improve the attack scheme that can fit Trojan in the limited space which reduces the effectiveness of the method. They showed that as the calculated activation value of clean input is not usually based on the Trojaned neurons, retraining the model with clean data does not correct the neural network adequately. Hence, they improved the method of \cite{68} by pruning prior to model training with malicious data. By doing so, the activation values of the benign and Trojaned inputs sometimes mapped to the same neurons. Consequently, retraining with clean input  modifies the neurons where Trojans reside. This removes  the Trojan in the neurons where the activations of both the benign and Trojaned samples occur by fine-tuning the model. 

Wu et al.~\cite{112} claimed that if neurons are maliciously perturbed, the neural network can easily  malfunction, categorising clean samples into the target class. They developed a neural Trojan defense method called Adversarial Neural Pruning (ANP). The ANP can help with the model correction by pruning the sensitive neurons while not significantly degrading the performance of the model. Zheng et al.~\cite{116} proposed a method that uses topological tools to model high-order dependencies in the neural network and detect the existence of neural Trojan. They found that there is an obvious difference in structures between the Trojaned and clean models where Trojaned model appear to have short-cuts going from the input to the output layers that do not exist in clean models. Therefore, by looking for the short-cuts in the neural network, they identify the Trojan.  

\subsection{Trigger-based Trojan Reversing}
Conceptually, trigger-based Trojan reversing estimates the potential trigger pattern for a model and uses it in model training/re-training to robustify the model against the trigger pattern.
Along this line, Wang et al.~\cite{70}  proposed a method termed Neural-Cleanse that works in three stages. First, it constructs potential triggers for each class and estimates a final synthetic trigger and target label.  Then, it  attempts to reverse the trigger effects based on model pruning and retraining. Lastly, it removes the Trojan by retraining network on the input with reverse engineered trigger to achieve model recovery. The authors  also demonstrate the vulnerability of deep models  against Trojan attack by examining the smallest size of perturbation required to cause model’s misbehaviour. Though effective, there is a limitation of Neural-Cleanse that it is incapable of dealing with Trojan of varied size, shape and location.

Guo et al.~\cite{71} proposed a method called TABOR that is able to overcome the limitation of Neural-Cleanse. TABOR utilises a non-convex optimization-theoretic formulation guided by explainable AI and other heuristics that enables the increase in the accuracy of detection without the restriction on trigger size, shape and location. Qiao et al.~\cite{72} also pointed out that the reversed  engineered triggers under~\cite{70} differ  significantly from the actual Trojan triggers. Inspired by this, they proposed a method to generalise the Trojan trigger. Instead of reversing all the individual triggers, they recovered the trigger distribution from the potential triggers to get a more precise reversed engineered trigger. The key idea of Qiao et al.~\cite{72} was agreed by Zhu et al.~\cite{74} who demonstrated the effectiveness of GAN-based synthesis of Trojan triggers  for model recovery.

Chen et al.~\cite{75} noticed the distinct difference in patterns of the neuron activation between  benign and Trojaned input in the final hidden layer of neural networks. Inspired by this, the authors proposed a detection method on the neuron activation pattern in the final hidden layer. The method involves forming clusters of the neuron activations in the final hidden layer and detecting Trojan by determining if abnormal characteristics are present in the cluster. The model can then be recovered by removing the clusters with abnormal characteristics and finetuing the model with clean input. Shen et al.~\cite{76} showed that Trojans can be removed by using only one class for trigger optimisation in each round of retraining. Aiken et al.~\cite{77} also proposed a method that combines model correction and trigger-based Trojan reversing involving pruning of neural network based on synthetic triggers. 

\subsection{Bypassing Neural Trojan}
This strategy involves using pre-processing to remove trigger in the input before passing the input to the model. Doan et al.~\cite{78} developed a technique, termed Februus to bypass Trojan triggers in images. Before the image enters the model, it is sent to Februus system to validate the presence of trigger, and remove it if is it suspected. Working of Februus to detects and neutralises the trigger can be understood as three steps. (i) Using a logit score-based method for Trojan detection that works under the assumption that if a trigger exists  in the input, then the predicted class is the target class. (ii) Remove the potential trigger with a  masking process. (iii) Use an inpainting technique to restore the image. Liu et al.~\cite{79} showed that an autoencoder can be used for image pre-processing to remove potential triggers. The autoencoder is placed between the image and the Trojaned model and it removes Trojan trigger by minimising the mean-squared error between the training set images and the reconstructed images.

Udeshi et al.~\cite{80} proposed a model-agnostic framework termed NEO to locate and mitigate Trojan trigger in the input images. NEO aims at predicting the correct outcomes of poisoned images and compares that  with the actual prediction. It places a trigger blocker on the images that has its prediction outcomes differing considerably from each other. In \cite{81}, Vasquez et al.~pre-processed images through the style transfer of the image. Li et al.~\cite{82} discovered that for the images with static trigger patterns, a slight change in the location or appearance of the trigger can  significantly degrade the effectiveness of Trojan attack. Inspired, they then proposed a method which transforms the input regularly by shrinking and flipping. Their technique is claimed to be an efficient detection method  that has low computational requirements. 

Zeng et al.~\cite{83} proposed a method that works by depressing the effectiveness of poisoned samples during the training time to prevent Trojan injection. The depression involves transformation of inputs in both training and run time process. Du et al.~\cite{84} used noisy stochastic gradient descent to learn the model. They demonstrated that when noise is present in the training set, the effectiveness of Trojan trigger reduces, resulting in a lower post-training attack success rate. Hong et al~\cite{85} took an alternative approach and observed that $l_2$ norm of the gradient of poisoned samples have significantly higher magnitude than benign samples and they also differ in their gradient orientation. They designed the differentially private stochastic gradient descent to perturb individual gradients of the training samples and trained the model with clean samples where all the Trojaned samples were removed from the training set. The existing neural Trojan detection methods often use an intermediate representation of models to distinguish between the Trojaned and benign models. Such methods are more effective  when spectral signature of the Trojaned data is sufficiently large. Hayase et al.~\cite{126}  proposed a robust covariance estimation method to amplify the spectral signature of the Trojaned data. %Therefore, a wide range of neural Trojan attacks can be detected and hence increase the chances for neural Trojan detection. 

\subsection{Input Filtering}
The input filtering strategy  involves filtering the malicious input so that the data passed to the model is likely clean. Works under this category can further be divided based on filtering applied at the training stage or the testing stage. 
%into training sample and testing sample filtering which filters the poisoning data from training and testing sample respectively. 
\subsubsection{Training Sample Filtering}
Tran et al.~\cite{86} noticed that there is a detectable trace for Trojaned samples in the spectrum of feature representation co-variance. Hence, they proposed to filter the Trojaned samples using the decomposition of  feature representation. Chen et al.~\cite{87} shared a similar idea to Tran et al.~\cite{86} and noted that the Trojaned and benign samples have different characteristic in the feature space. They demonstrated that the Trojaned samples in the training set can be filtered by first  performing clustering of training data neuron activation, then filtering the data by removing the cluster that represents poisoned samples. Tang et al.~\cite{88}  pointed out a limitation of the previous two methods, stating that simple target contaminations may result in  less distinguishable representations for benign and Trojaned samples. To overcome this, they proposed filtering based on representation decomposition and  statistical analysis of the individual samples. 
Similarly, Soremekun et al.~\cite{89} also proposed  filtering of poisoned samples based on the feature representation difference between the Trojaned and benign samples. Chou et al.~\cite{90} utilised saliency map for  detecting potential triggers in the input, and then they  filtered the samples containing the triggers. Li et al.~\cite{111} claimed that it is not evidental that there exist robust training methods to prevent the injection of triggers. They  conducted experiments which split the training process into clean data training and Trojaned data training. They found two weaknesses of Trojaned data training. 1) It is faster for the models to learn Trojaned data compared to the clean data and the time taken for the convergence of the model on the Trojaned data is highly dependent on the strength of the neural Trojan attack. 2) The neural Trojan always aims to lean the models to the Trojan target class. They leverage these observations to propose Anti-Backdoor Learning which can achieve neural Trojan prevention by isolating Trojaned samples at the training phase,  and weaken any potential relationship between the Trojaned sample and the target class.

% \begin{figure}
% \centerline{\includegraphics[width=20pc]{Images/RAID.png}}
% \caption{The Classification Accuracy (CA) and Attack Success Rate (ASR) of RAID against the number of updates. The top graph is a measure of CA and the bottom graph is measure of ASR. n is the number of Principal Component Analysis (PCA) components and the x-axis indicates the number of updates. It is obvious that the performance of RAID increases as the number of updates increases. Image taken from ~\cite{123}.}
% \label{RAID}
% \end{figure}

\begin{figure*}[t]
    \centering
     \includegraphics[width =\textwidth]{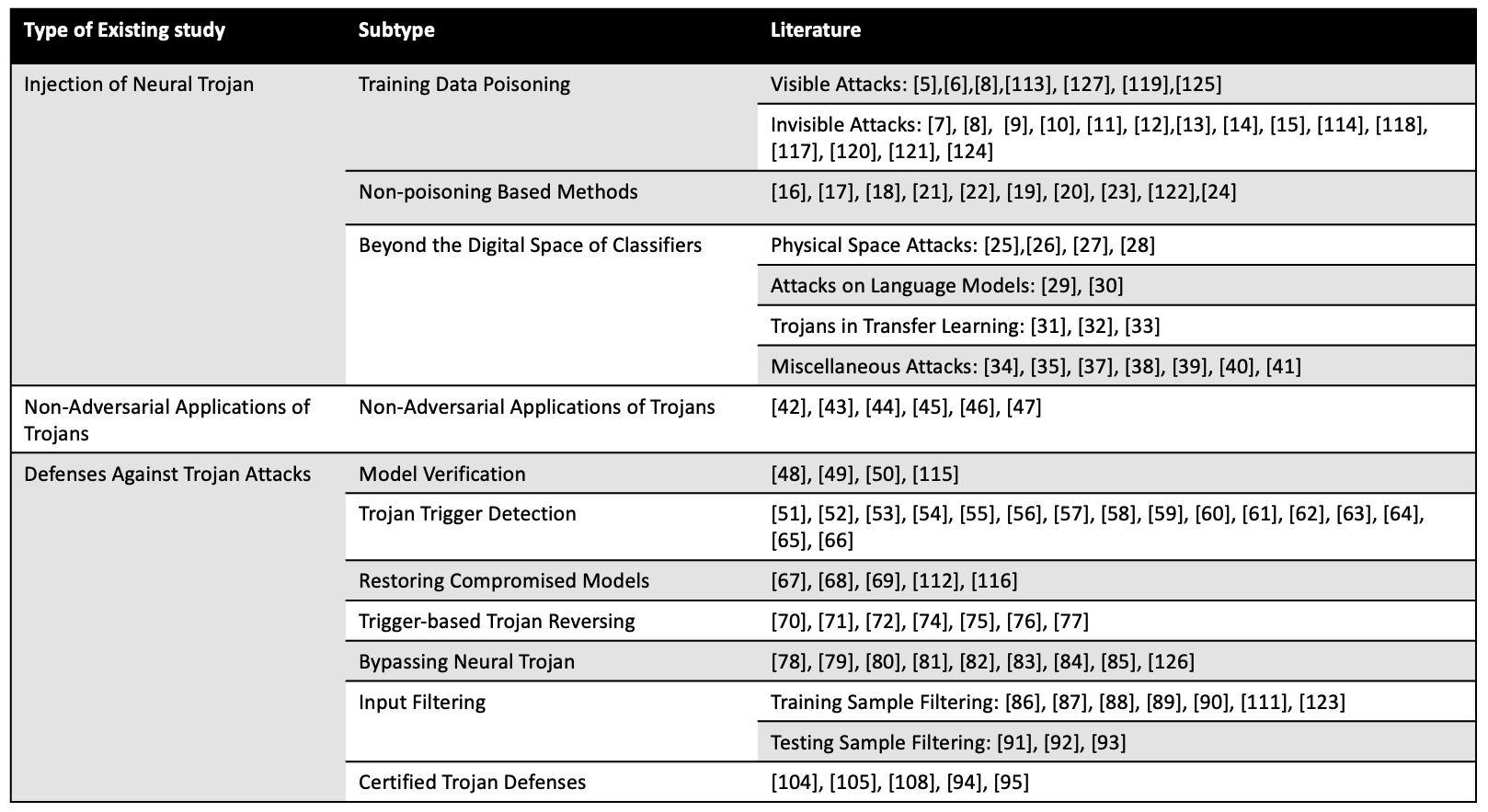}
    \caption{A summary of the prominent contemporary Neural Trojan injection and detection methods, along with non-adversarial applications of Neural Trojans. }
    \label{table}
\end{figure*}

Fu et al.~\cite{123} recently proposed a novel feature-based on-line detection strategy for neural Trojans that is named Removing Adversarial-Backdoors by Iterative Demarcation (RAID). This is achieved in two stages, which are off-line training and on-line retraining. The off-line training trains the neural network with only clean data first, and then the on-line retraining detects the input that is different from the clean data at the off-line training stage. Significantly different images are then removed. They next train a binary support vector machine (SVM) with both the purified anomalous data and the clean data so that RAID can use the SVM to detect the poisoned inputs in a dataset. The SVM is also designed to be updated in real-time.

\subsubsection{Testing Sample Filtering}
Similar to training sample filtering, the main goal of this type of detection method is to distinguish between the benign and Trojaned samples and filter the Trojaned ones from the entire set before passing them to the model. However, in this case, it is strictly done at the testing time only. 
%so that no Trojan trigger can be present for Trojan activation. The only difference is that this is conducted at the testing stage. 
Subedar et al.~\cite{91} proposed a method that is able to distinguish between the Trojaned and benign samples using model uncertainty during test time. Du et al.~\cite{92} demonstrated the effectiveness of outlier detection for the objective trigger detection while testing the model. Jaraheripi et al.~\cite{93} also proposed a lightweight method for sample filtering which does not require labeled data, model retraining or prior assumption on the design of the trigger and can work as testing stage filter.

\subsection{Certified Trojan Defenses}
Nearly all the above-mentioned defense techniques can be categorised as ad-hoc techniques that are based on heuristics. It is often mentioned in the literature that such defenses can be broken with the help of adaptive strategies~\cite{104}, \cite{105}. 
Hence, similar to the certified defenses for adversarial examples~\cite{108}, researchers have also started investigating such defenses for Trojan attacks. 
%In [134,135], Tan et al. and Saha et al. stated that almost all the aforementioned defense techniques fail to capture stronger adaptive attacks.
%Many researchers looked into this limitation and designed methods to overcome this. 
For instance, Wang et al.~\cite{94} proposed a random smoothing technique that adds random noise to the sample to ensure the robustness of the resulting model against the adaptive attacks. They improved the method by thinking the training procedure as base function and develop a smooth function based on the base function for smoothing. To an extent, Weber et al.~\cite{95} disagreed with the above and proved the ineffectiveness when applying smoothing directly. They proposed a framework that evaluates the difference in the smoothing noise distributions to achieve better robustness of the model.

\section{Discussions and Future Outlook}

This survey focused on the literature published from 2017 to 2021 on neural Trojan with nearly 60\% of the paper on Trojan attacks and the other 40\% on defenses against the attacks. Incidentally, we notice more activity in neural Trojan literature since 2019, as compared to previous years. This indicates an increasing interest of the research community in this direction.  This trend is inline with the literature in a closely related research direction of adversarial attacks on deep learning~\cite{108}. We conjecture that this ever increasing research activity in these directions is a natural consequence of awareness of vulnerabilities of deep learning in adversarial setups.
We noticed a clear cat-and-mouse game between Trojan injections and Trojan defenses until the Triggerless Backdoor~\cite{24} and dynamic backdoor~\cite{25} appeared in the literature. These attacks have been shown to bypass the state-of-the-art defense technologies at their time, which shows that the game is being led by the attack methods. This battle is still on though, resulting in further discoveries of vulnerabilites of deep learning and their remedies. 

Whereas the literature discussed in the preceding sections cover a wide range of topics and possibilities,
this research direction is relatively new. Hence, there is still considerable opportunity to explore new sub-topics in this direction. A guide to such an exploration is provided by the sister problem of adversarial attacks on deep learning~\cite{108}. The discovery of adversarial attacks was made in 2013 - a few year earlier than identification of neural Trojans. Hence, that problem is currently relatively more popular in the literature. The maturity of literature in adversarial attacks can provide useful guideline for research in Trojan attacks. We also list a few possible future directions and challenges for Trojan attacks based on our literature review and other related problems.

%discussed above addressed the researches conducted in discovering the Trojan injection and defenses techniques. However, from those techniques, we can still see that there are areas that left uncovered or where more attentions are required. Listed below are some potential research areas that can be further investigated in the future.

\subsection{Focus on Black-box Attacks}
Almost 95\% of the currently available Trojan attacks  are white-box or grey-box that are based on the assumption that at least  parts of the training data is accessible to the attackers. For example, all the attacks introduced by  training data poisoning assume that complete training data is available to insert the  backdoor in the target model. In real-life, full training data is usually not accessible due to privacy reasons. Hence, these white-box and grey-box setups do not fully apply to practical scenarios. This makes exploration of methods to embed Trojans or exploit triggers under fully black-box setup an interesting future direction. It may first seem that black-box scenario does not apply to Trojan attacks because Trojans are embedded in the model itself, and embedding them must require model or training data access. However, it may be possible to identify natural vulnerabilities of models by querying them, e.g.~discovering sensitivity of a model to irrelevant pattern. This can subsequently be exploited in embedding triggers in input during test time.  
%needed for designing black-box attacks where there is no access to the training data available.

\subsection{Attacks and Defenses Beyond Visual Models}
Deep learning has its applications far beyond visual models. Currently, a wide range of application tasks and neural network types are exploiting deep learning, e.g.~in speech recognition~\cite{29}, \cite{30}, graph networks~\cite{34}, \cite{35} etc. It can be notice in our literature review that an overwhelming majority of the existing works are concerned with visual models only. In other words, most of the Trojan embedding techniques are tailored to convolutional neural networks operating in the domain of images.  %of the studies are focused on using images as inputs while 40\% of the papers concentrating on all other fields. 
In this case, the effectiveness of the attacks can be improved by maximising the attack success rate while keeping the Trojan trigger hidden by blending the noise throughout the images. However, designing an invisible trigger pattern in the other domains, e.g.~speech recognition, natural language processing, can be significantly different as the triggers can no longer be blended into e.g.~sentences. 
%Therefore, more researchers can look into the optimal method to insert Trojan triggers in those areas.
This opens new challenges for research in Trojan attacks in different domains. We anticipate that in the future, this direction will still encounter uncharted territories when its scope will be expanded to other domains. Indeed, it is obvious that those domains will be found susceptible to Trojan attacks as long as they exploit the technology of deep learning.  

\subsection{Efficacy of  Trojan Design}
Many works in the related literature focus on finding the most effective and secretive way to insert Trojan triggers into the inputs. Nonetheless, the actual pattern  of the Trojan triggers are equally important, whose efficacy is relatively less explored in the current literature.  The c-BaN in dynamic backdoor~\cite{15}  is one of the few methods to generate triggers that are most suited to  images with a given label. Others generally  think of Trojan triggers as a single pattern or use simple algorithms for trigger generation. Hence, the ways to design powerful trigger patterns can still be explored. It is also mentioned in \cite{4} that most of the studies only consider the effectiveness and invisibility of triggers. However, more research can be conducted, aiming to design a Trojan trigger that requires a minimised amount of training data to be poisoned.

\subsection{Stronger Defenses for the Existing Attacks}
As mentioned earlier, the recent triggerless backdoors~\cite{24} and dynamic backdoors~\cite{15} are designed to be too effective to be detected by the state-of-the-art defense techniques. Triggerless backdoor broke the traditional way of thinking and enabled Trojan activation without the presence of a trigger. Dynamic backdoor also improved on the design of traditional Trojan to allow the trigger to be a random pattern  at random locations.  This makes the detection of such a trigger impossible. With continuous developments of Trojan embedding techniques and introduction of more and more powerful attacks, we expect to see a counter stream of stronger defenses in the future. %Thus, more powerful defenses need to be discovered to capture these attacks.

\section{Conclusion}
Neural Trojan is a serious problem for deep learning technology as it affects neural networks such that they work normally for benign inputs, but maliciously in the presence of a trigger in the input.
%whereas when there is a Trojan pattern present, it will trigger the neural network's malicious behaviour and output a target-specific label. 
This has serious implications for security-critical applications. 
For example, in facial recognition, an attacker can exploit neural Trojan  to grant authentication to irrelevant personnel to sensitive areas or information. Detection of such an attack is not easy because the model operates normally in most cases and misbehave only in very specific cases. 
%for people who do not have access without user's realisation as the deep learning model works correctly in all other situations. 
Due to the large growth of the applications of deep learning, Trojan attacks have caught attentions of many researchers in the recent years. This has also caused defenses against Trojan attacks to emerge. 
%have become more and more critical for the security of the deep learning model. 
We noticed that every year, there is an increasing number of works appearing in the reputed sources of machine learning and computer vision, e.g.~CVPR, ICCV, ECCV, ICLR, NeurIPS that are concerned with neural Trojans. This observation leads us to believe that this research direction is likely to become even more popular in the near future. 
%, looking into this and with more and more powerful attacks and defenses proposed. 
Whereas there have already been a few literature reviews in this direction, our survey is unique in that it we reviews papers that are published recently. It  summarised the recent neural Trojan injection and defense techniques by systematic categorization and discussed their effectiveness. 
%The first section briefly introduces deep learning including its applications, how deep learning works as well as why is there a potential risks of Trojan injection in deep learning model. The second section divided the methods for Trojan injection into 3 categories, training data poisoning, non-poisoning based methods and Trojan injection in other spaces or fields. The third section elaborated on how Trojan can be used for good. The fourth section discussed the Trojan defense techniques by neural network verification, Trojan trigger detection, restoring compromised neural network, trigger-based Trojan reversing, bypassing neural Trojan, input filtering and certified Trojan defenses. The future challenges are also listed for future investigation at the end.

\section*{References}

\def\refname{}

\end{document}